\documentclass[universe,article,accept,moreauthors,pdftex,10pt,a4paper]{mdpi}

\usepackage{bm}
\usepackage{amsmath}
\usepackage{graphicx}
\usepackage{diagbox}
\usepackage[usenames,dvipsnames]{color}
\definecolor{darkblue}{RGB}{0,0,192}
\definecolor{darkgreen}{RGB}{0,122,22}

\usepackage{booktabs}
\usepackage{multirow}
\usepackage{soul}
\usepackage{microtype}

\firstpage{1}
\articlenumber{x}
\doinum{10.3390/------}
\pubvolume{2}
\pubyear{2016}
\copyrightyear{2016}
\externaleditor{Academic Editor: {name}}
\history{Received: 5 August 2016; Accepted: 17 August 2016; Published: {date}}

\Title{Predictions for Bottomonia Suppression in 5.023 TeV Pb-Pb Collisions} 

\Author{{Brandon} Krouppa and Michael Strickland *} 

\AuthorNames{Brandon Krouppa and Michael Strickland }

\address [1]{Department of Physics, Kent State University, Kent, OH 44242,  USA; {bkrouppa@kent.edu}}

\corres{Correspondence: mstrick6@kent.edu}

\abstract{We compute the  suppression of the bottomonia states $\Upsilon(1S)$, $\Upsilon(2S)$, $\Upsilon(3S)$, $\chi_{b}(1P)$, $\chi_{b}(2P)$, and $\chi_{b}(3P)$ states in {Large Hadron Collider (LHC)} $\sqrt{s_{NN}} = 5.023$ TeV Pb-Pb collisions.  For the background evolution we use 3+1d anisotropic hydrodynamics with conditions extrapolated from $\sqrt{s_{NN}} = 2.76$ TeV and we self-consistently compute bottomonia decay rates including non-equilibrium corrections to the interaction potential.   For our final results, we make predictions for $R_{AA}$ as function of centrality, rapidity, and $p_{T}$ for the $\Upsilon(1S)$ and $\Upsilon(2S)$ states, including feed down effects.  In order to assess the dependence on some of the model assumptions, we vary the shear viscosity-to-entropy density ratio, $4\pi\eta/s \in \{1,2,3\}$, and the initial momentum-space anisotropy parameter, $\xi_{0} \in \{0, 10, 50\}$, while holding the total light hadron multiplicity fixed.
}


\begin{document}
\section {Introduction}
\label{sec:intro}

Ultra-relativistic heavy ion collisions (URHICs) carried out at the Large Hadron Collider (LHC) at CERN and the Relativistic Heavy Ion Collider (RHIC) at Brookhaven National Laboratory create super-hot and dense matter.   The observed thermal particle spectra and anisotropic collective flow seen in such collisions has been taken as evidence of the creation of a deconfined high-temperature quark-gluon plasma (QGP).  The existence of the QGP itself is a prediction of finite-temperature quantum chromodynamics (QCD), and the QGP is believed to have been the state of the early universe.

Comparisons between theory and experiment suggest that LHC URHICs produce a QGP with an initial temperature on the order of $T_{0}=500{-}600$ MeV for \mbox{$\sqrt{s_{NN}} = 2.76$ TeV} Pb-Pb collisions~\cite{Heinz:2013th,Gale:2013da}. The LHC has recently performed URHICs with nucleon--nucleon center-of-mass energies of \mbox{$\sqrt{s_{NN}} = 5.023$ TeV}, which are expected to produce initial temperatures on the order of $T_{0}=600{-}700$~MeV.  Analysis of collective flow data indicate that the QGP behaves like a nearly-perfect fluid with shear viscosity-to-entropy density ratio $\eta/s$ approaching the conjectured lower bound of $1/(4\pi)$. Despite being nearly ideal, one complication that must be faced is that, although the QGP is close to thermal equilibrium at late times, URHICs produce a QGP which is momentum-space anisotropic in the local rest frame with the degree of momentum-space anisotropy decreasing slowly as a function of proper time.  This can have an impact on the various signatures for QGP formation, such as heavy quarkonium suppression \cite{Strickland:2014pga,Fukushima:2016xgg}.

Heavy quark-bound states are of particular interest because they can survive into the deconfined phase due to their large binding energy.  They are expected to survive up to a few times the pseudo-critical temperature, $T_{c}$; however, at sufficiently high temperatures, even heavy quark bound states should disassociate.  As a result, in the late 1980's Karsch, Matsui, Mehr, and Satz predicted that heavy quark--antiquark bound states, known as quarkonia, would be suppressed in heavy ion collisions compared to proton--proton (p-p) collisions \cite{Matsui:1986dk,Karsch:1987pv}.   The resulting suppression of heavy quarkonia can be used to probe the entire history of the QGP, since heavy quarkonium states are produced early in the QGP evolution.
There has been a significant amount of work dedicated to the study of heavy quarkonia suppression.  For recent related work, see References~\cite{Emerick:2011xu,Nendzig:2012cu,Zhou:2014kka,Hoelck:2016tqf}.   In this paper, we focus on  bottomonia states which survive up to plasma temperatures of $T_d \sim 600, 230,$ and 170 MeV for $\Upsilon(1S)$, $\Upsilon(2S)$, and $\Upsilon(3S)$ states, respectively \cite{Mocsy:2013syh}. At these temperatures, the in-medium width of the state becomes on the order of the real part of its binding energy, and the state quickly dissociates; however, prior to this disassociation point, the states have large widths ($\sim$50--100 MeV), with the widths for the excited states being larger than those for the ground state.   As a consequence, one expects sequential melting of heavy bottomonia states in QGP, with excited states melting before the ground state, and lighter states dissociating before heavier states \cite{Karsch:2005nk}.

In this paper, we focus on bottomonia states since they can be more reliably treated using potential-based non-relativistic effective field theory (pNRQCD) \cite{PhysRevD.21.203,Lucha:1991vn,Brambilla:2004jw,Brambilla:2008cx,Brambilla:2010xn}.  We include the effect of in-medium disassociation in the presence of a complex-valued potential obtained using hard-thermal-loop perturbation theory~\cite{Laine:2006ns,Dumitru:2007hy,Burnier:2009yu,Dumitru:2009fy,Dumitru:2009ni}.  We also take into account the non-equilibrium momentum-space anisotropy produced as a result of the rapid longitudinal expansion of the QGP via modifications to both the real and imaginary parts of the potential~\cite{Margotta:2011ta}.  This work extends our previous work~\cite{Krouppa:2015yoa,Strickland:2011aa,Strickland:2011mw} to \mbox{$\sqrt{s_{NN}} = 5.023$ TeV} Pb-Pb collisions.  As before, we use 3 + 1d anisotropic hydrodynamics (aHydro) for the background evolution~\cite{Florkowski:2010cf,Martinez:2010sc,Bazow:2013ifa,Strickland:2014pga,Nopoush:2015yga,Bazow:2015cha}.

For the $\sqrt{s_{NN}} = 5.023$ TeV initial conditions, we extrapolate the initial conditions used at lower collision energies and make predictions for the inclusive $R_{AA}$ for the $\Upsilon(1S)$ and $\Upsilon(2S)$ states, including feed down contributions.  To assess the dependence on some of the model assumptions, we vary the shear viscosity-to-entropy density ratio, $4\pi\eta/s \in \{1,2,3\}$, and the initial momentum-space anisotropy parameter, $\xi_{0} \in \{0, 10, 50\}$, while holding the total light hadron multiplicity fixed.

\section {3 + 1\lowercase{d} Anisotropic Hydrodynamics}
\label{sec:31d}

In aHydro, the local rest frame (LRF) partonic distribution is assumed to be of the form~\cite{Martinez:2010sc,Florkowski:2010cf,Romatschke:2003ms}
\begin{equation}
 \label{eq:dist}
f(p,x) = f_{\text{eq}}\!\left(\sqrt{p_{T}^{2}+[1+\xi(x)]p_{z}^{2}}/\Lambda(x)\right),
\end{equation}
where $-1 \leq \xi(x) < \infty$ is the local momentum-space anisotropy parameter, $\Lambda(x)$ is the local transverse temperature of the plasma, and $f_{\text{eq}}$ is an arbitrary isotropic equilibrium distribution function which here we take to be a Boltzmann distribution.  The local anisotropy $\xi(x)$ encodes the viscous corrections to an ideal isotropic QGP.  Note that, in the most recent applications of aHydro, one allows for the existence of three independent anisotropy parameters; see e.g., Reference~\cite{Nopoush:2015yga}.  This takes into account both bulk and shear viscous effects in the QGP.  In this paper, we use the simpler ``one anisotropy parameter''~version.  This is sufficient for our purposes, since the additional anisotropy parameters are small and, therefore,~subleading.
  For details concerning the aHydro dynamical equations and the numerical methods used, see References~\cite{Martinez:2012tu,Ryblewski:2012rr}.

We start the hydrodynamic evolution of the system at $\tau_0 =$ 0.3 fm/c.  For the initial conditions, we use the Glauber model with a linear combination of wounded-nucleon and binary collisions.  The~fraction of the binary collision  component is taken to be $\kappa_{\text{binary}} = 0.145$.  The inelastic cross-section for $\sqrt{s_{NN}} = 2.76$ TeV collisions is taken to be $\sigma_{NN}^{2.76 \text{ TeV}} = 62$ mb, while for $\sqrt{s_{NN}} = 5.023$ TeV collisions, we extrapolate from the results presented in Reference \cite{Miller:2007ri} to obtain $\sigma_{NN}^{5.023 \text{ TeV}} = 67$ mb.   In both case,s we use a standard Woods--Saxon profile for the incoming nuclei with \mbox{$n_{A}(r) = n_{0}/(1 + e^{(r-R)/d})$}, where $n_{0} = 0.17 \ \text{fm}^{-3}$ is the central nucleon density and is determined via the normalization $\lim_{A\rightarrow \infty} \int d^{3}r \ n_{A}(r) = A$, $R = (1.12A^{1/3} - 0.86A^{-1/3})$ fm is the nuclear radius, and $d = 0.54$ fm is the skin depth of the nucleus \cite{florkowski2010}.

The initial longitudinal profile is approximated by distribution featuring a nearly boost-invariant plateau region for mid-rapidities and a tail characterized by a Gaussian which reflects limiting fragmentation~\cite{Bozek:2012qs}
\begin{equation}
{\cal F}(\varsigma) \equiv \exp \left[-\dfrac{(\varsigma - \Delta\varsigma)^{2}}{2\sigma_{\varsigma}^{2}}\Theta(|\varsigma|-\Delta\varsigma)\right].
\end{equation}

For $\sqrt{s_{NN}} = 2.76$ TeV collisions, fits were made to experimental particle pseudorapidity profiles which give ${\Delta\varsigma}^{2.76 \text{ TeV}} = 2.5$ and $\sigma_{\varsigma}^{2.76 \text{ TeV}}$ = 1.4. For $\sqrt{s_{NN}} = 5.023$ TeV collisions, we use the predictions of Reference \cite{Ozonder:2013moa}, which determined the rapidity profile in the high energy limit. Using the results of Reference \cite{Ozonder:2013moa}, one finds that going from $2.76$ TeV collisions to $5.023$ TeV collisions, there is an approximately $12\%$ increase in plateau halfwidth, which gives ${\Delta\varsigma}^{5.023 \text{ TeV}} = 2.8$. Based on \cite{Ozonder:2013moa}, we found no observable change of the Gaussian halfwidth in $5.023$ TeV collisions,  and therefore we take $\sigma_{\varsigma}^{5.023 \text{ TeV}} = 1.4$.  The combination of Glauber in the transverse plane and the longitudinal profile function above gives us the full 3d initial energy density profile for the QGP.

The initial plasma temperatures, $T_{0}$, for $2.76$ TeV collisions are taken from our previous work~\cite{Krouppa:2015yoa} and are consistent with the analysis of elliptic flow coefficients \cite{Schenke:2011tv}.  In Table \ref{table1}, we show the initial conditions used at 2.76 TeV. The values in the center of Tables \ref{table1} and \ref{table2} are the initial transverse temperatures of the plasma, which enter into the one-particle distribution function, Equation (\ref{eq:dist}). Note~that unless $\xi_{0}=0$, the transverse temperature cannot be interpreted as temperature. The entry with $\xi_{0}=0$ and $4\pi\eta/s = 1$ was obtained via fits to collective flow, and the other entries were obtained by varying either $\xi_0$ or $\eta/s$ while holding the total particle multiplicity fixed.  Note that, when $\xi_0 \neq 0$, we set the initial condition for the transverse temperature $\Lambda_0$ instead of the initial temperature due to the momentum-space~anisotropy.

\begin{table}[H]
\centering

\caption{$\Lambda_{0}$ values for $2.76$ TeV collisions in GeV.}
\label{table1}
\begin{tabular}{cccc}
\toprule
\backslashbox{\boldsymbol {$4\pi\eta/s$}}{\boldsymbol {$\xi_{0}$}} & \textbf{0} & \textbf{10} & \textbf{50}\\
\midrule
1 & 0.552 & 0.765 & 0.925\\
2 & 0.546 & 0.752 & 0.909\\
3 & 0.544 & 0.748 & 0.906\\
\bottomrule

\end{tabular}

\end{table}

For $5.023$ TeV collisions, we assume that the temperature of the plasma scales proportionally with the fourth root of the collision energy---i.e., $T_{0} \propto s_{NN}^{1/8}$. Using this assumption, we predict a $16\%$ increase in the initial central temperature when going from 2.76 TeV  to \mbox{5.023 TeV} collisions.  We apply this scaling to the \mbox{2.76 TeV} initial conditions with $\xi_{0}=0$ and $4\pi\eta/s = 1$ and then, as before, we fill out the rest of the initial conditions table by holding the total particle multiplicity fixed.  The resulting table of initial conditions is shown in Table \ref{table2}.

\begin{table}[H]
\centering
\caption{$\Lambda_{0}$ values for $5.023$ TeV collisions in GeV.}
\label{table2}
\begin{tabular}{cccc}
\toprule
\backslashbox{\boldsymbol {$4\pi\eta/s$}}{\boldsymbol {$\xi_{0}$}} & \textbf{0} & \textbf{10} & \textbf{50}\\
\midrule
1 & 0.641 & 0.888 & 1.076\\
2 & 0.632 & 0.869 & 1.053\\
3 & 0.629 & 0.863 & 1.046\\
\bottomrule
\end{tabular}
\end{table}

\section {Model Potential}
\label{sec:potential}

Modeling bottomonia states in a QGP which is anisotropic in momentum-space requires going beyond the Karsch, Matsui, Mehr, and Satz model which describes the free energy of a static heavy quark--antiquark pair in an isotropic plasma.  It has been shown in References~\cite{Emerick:2011xu, Strickland:2011aa} that free-energy-based potential models of the quarkonium pair do not agree with experimental $R_{AA}$ data.  For this reason, we use the internal energy of the bottomonia pair $U = F + T S$ to provide the model potential. The masses of heavy quarks allow bottomonia pairs to be treated using pNRQCD methods which allow us to systematically include relativistic corrections. We use a model which represents the short- and medium-range gluonic screening of the heavy-quark potential in a momentum-space anisotropic plasma~\cite{Strickland:2011mw,Strickland:2011aa,Dumitru:2007hy,Dumitru:2009ni,Dumitru:2009fy}.   For the long range real part of the potential, we use the original Karsh--Mehr--Satz (KMS) form for the free energy.  The resulting internal-energy-based model has the following form for the real part of the heavy quark potential

\begin{equation}
\Re[V] = -\dfrac{a}{r}(1 + \mu r)e^{-\mu r} + \dfrac{2\sigma}{\mu}[1-e^{-\mu r}] - \sigma r e^{-\mu r} - \dfrac{0.8\sigma}{m_{b}^{2}r},
\end{equation}

where $\mu = \mathcal{G}(\xi,\theta)m_{D}$ \cite{Strickland:2011aa,Dumitru:2007hy,Dumitru:2009ni}, with $\theta$ being the angle between the line connecting the quark--antiquark pair and the beam-line direction, $m_{D}$ is the isotropic leading-order Debye mass, $a = 0.385$, $\sigma = 0.223 \ \text{GeV}^{2}$ \cite{Petreczky:2010yn}, and $m_{b} = 4.7$ GeV is the mass of the bottom quark. The last term taken from Reference \cite{Bali:1997am} accounts for the temperature- and spin-independent finite-quark-mass correction, to obtain our final complex-valued potential (This term is not particularly important for bottomonia states.  We include it for historical continuity with our previous works, where we also considered charmonium suppression). For the imaginary part of the potential, we use a small-$\xi$ expansion of the heavy-quark potential~\cite{Laine:2006ns,Dumitru:2009fy,Burnier:2009yu}

\begin{equation}
\Im[V] = -\alpha_{s}C_{F}T\Big\{\phi(\hat{r}) - \xi[\psi_{1}(\hat{r},\theta)+\psi_{2}(\hat{r},\theta)]\Big\} ,
\end{equation}
where $\hat{r}=rm_{D}$, $\phi(\hat{r})$ is defined as
\begin{equation}
\phi(\hat{r}) = 2\int_{0}^{\infty} dz \dfrac{z}{(z^2+1)^2}\left[1-\dfrac{\sin\left(z\hat{r}\right)}{\hat{r}}\right],
\end{equation}
and $\psi_{1}$ and $\psi_{2}$ are defined as follows

\begin{equation}
 \psi_1(\hat{r}, \theta) = \int_0^{\infty} dz
 \frac{z}{(z^2+1)^2}\left(1-\frac{3}{2}
 \left[\sin^2\theta\frac{\sin(z\, \hat{r})}{z\, \hat{r}}
 +(1-3\cos^2\theta)G(\hat{r}, z)\right]\right),
 \end{equation}

 \begin{equation}
 \psi_2(\hat{r}, \theta) = - \int_0^{\infty} dz
\frac{\frac{4}{3}z}{(z^2+1)^3}\left(1-3 \left[
  \left(\frac{2}{3}-\cos^2\theta \right) \frac
 {\sin(z\, \hat{r})}{z\, \hat{r}}+(1-3\cos^2\theta)
 G(\hat{r},z)\right]\right).
\label{eq:psis}
\end{equation}

The algorithm from Reference \cite{Strickland:2009ft,Margotta:2011ta} is used to solve the resulting 3d Schr\"odinger equation on a regular lattice by transforming to imaginary time and using the finite difference time domain (FDTD) method.   Using this method, we compute the real and imaginary parts of the binding energy over a range of $\Lambda$ from 144 MeV to 1037 MeV. For each $\Lambda$, we compute the real and imaginary binding energies for a range of anisotropies, $\xi$, from $-0.3$ to $200$, with an irregular spacing which accounts for the fact that the majority of the time the system probes small values of $\xi$.  We use a $N^{3} = 256^{3}$ lattice with lattice spacing $a = 0.1 \ \text{GeV}^{-1} \approx 0.02$ fm and $a = 0.15 \ \text{GeV}^{-1} \approx 0.03$ fm for a total box length of $L = Na \approx 5.04$ fm and $L = Na \approx 7.56$ fm, for $\Upsilon$ and $\chi_{b}$ states, respectively. The imaginary-time step size for the algorithm is taken to be $\Delta\tau = a^{2}/8$.

The real and imaginary binding energies are extracted using~\cite{Strickland:2011aa}
\begin{equation}
E_{\nu,\text{bind}} \equiv -\left(E_{\nu}-m_{1}-m_{2}-\dfrac{\langle\phi_{\nu}|V_{\infty}(\theta)|\phi_{\nu}\rangle}{\langle\phi_{\nu}|\phi_{\nu}\rangle}\right) ,
\end{equation}
where
\begin{equation}
V_{\infty}(\theta) \equiv \lim_{|{\bf r}|\rightarrow\infty} \Re[V(\theta,{\bf r})] \, .
\end{equation}
%

Negative values of $\Im[E_{\text{bind}}]$ only occur for large values of $\xi$, which is a consequence of the small-$\xi$ expansion. Large values of $\xi$ correspond to a nearly free streaming quark-gluon plasma, so it is expected that the widths of bottomonia states return to vacuum values, which are on the order of $\sim$~keV, which effectively allows us to set $\Im[E_{\text{bind}}] = 0$ for this specific case.

\section {{Bottomonium} Suppression} 
\label{sec:raa}

We use the following rate equation to account for in-medium bottomonia state decay,
\begin{equation} \label{eq:rate}
\dfrac{dn(\tau,{\bf x}_\perp,\varsigma)}{d\tau} = -\Gamma(\tau,{\bf x}_\perp,\varsigma)n(\tau,{\bf x}_\perp,\varsigma) ,
\end{equation}
where all variables depend on longitudinal proper time $\tau = \sqrt{t^{2} - z^{2}}$, the transverse coordinate ${\bf x}_{\perp}$, and the spatial rapidity $\varsigma = {\rm arctanh}(z/t)$. The rate of decay is computed by~\cite{Strickland:2011aa}
\begin{equation}
\Gamma(\tau, {\bf x}_{\perp}, \varsigma) = 
\begin{cases} 
2\Im[E_{\text{bind}}(\tau, {\bf x}_{\perp}, \varsigma)] & \Re[E_{\text{bind}}(\tau, {\bf x}_{\perp}, \varsigma)] > 0 \\ 
\gamma_{\text{dis}} & \Re[E_{\text{bind}}(\tau, {\bf x}_{\perp}, \varsigma)] \leq 0. 
\end{cases}
\end{equation}
where $\gamma_{\text{dis}}$ is a large value which is chosen such that a completely unbound state decays quickly.  We~emphasize that $E_{\text{bind}}$ and hence $\Gamma$ are local quantities of $\tau$, ${\bf x}_{\perp}$, and $\varsigma$ through the 3 + 1d background evolution of the transverse temperature $\Lambda(x)$ and the local momentum-space anisotropy $\xi(x)$.

\subsection{Survival Probability}
\label{ssec:survival}

The survival probability is determined by integrating Equation~(\ref{eq:rate}) over proper time with the integration limits set dynamically. The lower integration limit for the proper-time integration is ${\rm max}(\tau_{\text{form}},\tau_{0})$, where $\tau_{0}$ is the initial proper time for plasma evolution and $\tau_{\text{form}}$ is the time-dilated formation time of the state in question, computed via $\tau_{\text{form}}(p_{T}) = \gamma \tau_{\text{form}}^{0} = E_{T}\tau_{\text{form}}^{0}/M$, where $M$ is the mass of the state.  The rest-frame formation time $\tau_{\text{form}}^{0}$ for each of the states is taken to be inversely proportional to the vacuum binding energy of each state \cite{Karsch:1987uk}, and thus, $\tau_{\text{form}}^{0} =$ 0.2, 0.4, 0.6, 0.4, 0.6, and 0.6 fm/c, for $\Upsilon(1S)$, $\Upsilon(2S)$, $\Upsilon(3S)$, $\chi_{b}(1P)$, $\chi_{b}(2P)$, and $\chi_{b}(3P)$, respectively. The upper limit for the proper-time integration is determined to be the proper time at which the local energy density becomes less than the energy density of an $N_{c} = 3$ and $N_{f} = 2$ ideal gas of quarks and gluons with a temperature of $T_f = 192$ MeV. This is the temperature at which screening effects are assumed to turn off rapidly due to the transition to the hadronic phase.

Transverse momentum cuts were implemented by assuming that the transverse momentum distribution of bottomonia states is proportional to $E_{T}^{-4}$,
\begin{equation}
R_{AA}({\bf x}_{\perp},\varsigma) \equiv \dfrac{\int_{p_{T,\text{min}}}^{p_{T,\text{max}}} dp_{T}^{2} \, R_{AA}(p_{T},{\bf x}_{\perp},\varsigma)/(p_{T}^{2}+M^{2})^{2}}{\int_{p_{T,\text{min}}}^{p_{T,\text{max}}} dp_{T}^{2}/(p_{T}^{2}+M^{2})^{2}}.
\end{equation}
Once the transverse momentum cut is applied, we average $R_{AA}$ over the transverse plane
\begin{equation}
\langle R_{AA} (\varsigma) \rangle \equiv \dfrac{\int_{{\bf x}_{\perp}} n_{AA}({\bf x}_{\perp})R_{AA}({\bf x}_{\perp},\varsigma)}{\int_{{\bf x}_{\perp}} n_{AA}({\bf x}_{\perp})} \, ,
\end{equation}
where $T_A(x,y) = \int_{-\infty}^{\infty} dz \ n_{A}\left(\sqrt{x^2 + y^2 + z^2}\right)$ is the nuclear thickness function and $n_{AB} = T_{A}(x+b/2,y)T_{B}(x-b/2,y)$ is the overlap density function. In the above relations, it is assumed that $n_{AA}$ sets the probability for bottomonia production at a given point in the transverse plane. After the $p_T$ cuts and spatial averaging are performed, centrality averaging is performed by converting the impact parameter, $b$, to centrality, $C$, using the Glauber model, and then integrating over the centrality cut with a probability distribution proportional to $e^{-C/20}$ with $0 < C < 100$ \cite{Chatrchyan:2012np}.

\subsection{Excited State Feed Down}
\label{ssec:feeddown}

A certain fraction of $\Upsilon(1S)$ and $\Upsilon(2S)$ states produced in URHICs are formed via the decay of excited states. To compute the post feed down $R_{AA}$ for the $\Upsilon(1S)$ and $\Upsilon(2S)$ states, $R_{AA}$ is computed taking into account the suppression of excited states which undergo late time feed down.  For this purpose, we use $p_{T}$-averaged feed down fractions obtained recently from a compilation of p-p data available from ATLAS, CMS, and LHCb 
 \cite{Woeri:2015hq}.  The fractions $f_{i}^{\Upsilon(1s)}$ and $f_{i}^{\Upsilon(2s)}$ are given in \mbox{Tables \ref{feeddown1} and \ref{feeddown2}}, which specify the fraction of a particular state which contributes to the $\Upsilon(1S)$ and $\Upsilon(2S)$ states. The~inclusive feed down $R_{AA}$ of each of the states is calculated using a linear combination of the primordial $R_{AA}$ using the respective feed down fractions $R_{AA}^{\Upsilon(nS)} = \sum_{i \  \in \  \text{states}} f_{i}^{\Upsilon(nS)}R_{AA,i}$, where $R_{AA,i}$ is the primordial suppression for the $i$th  state.
\begin{table}[H]
\centering
\small
\begin{tabular}{cc}
\toprule

\multicolumn{2}{c}{\boldsymbol{$\Upsilon(1S)$}\textbf{ Feed Down Fractions}}\\
\midrule

$\Upsilon(1S)$ & 0.668\\
$\Upsilon(2S)$ & 0.086\\
$\Upsilon(3S)$ & 0.010\\
$\chi_{b}(1P)$ & 0.170\\
$\chi_{b}(2P)$ & 0.051\\

$\chi_{b}(3P)$ & 0.015\\
\bottomrule

\end{tabular}
\caption{Feed down fractions to the $\Upsilon(1S)$ state.}
\label{feeddown1}
\end{table}

\begin{table}[H]
\centering
\begin{tabular}{cc}
\toprule
\multicolumn{2}{c}{\boldsymbol{$\Upsilon(2S)$} \textbf{Feed Down Fractions}}\\
\midrule
$\Upsilon(2S)$ & 0.604\\
$\Upsilon(3S)$ & 0.043\\
$\chi_{b}(2P)$ & 0.309\\
$\chi_{b}(3P)$ & 0.044\\
\bottomrule
\end{tabular}
\caption{Feed down fractions to the $\Upsilon(2S)$ state.}
\label{feeddown2}
\end{table}

\subsection{Results}
\label{ssec:results}

\begin{figure}[H]
\centerline{
\includegraphics[width=0.47\linewidth]{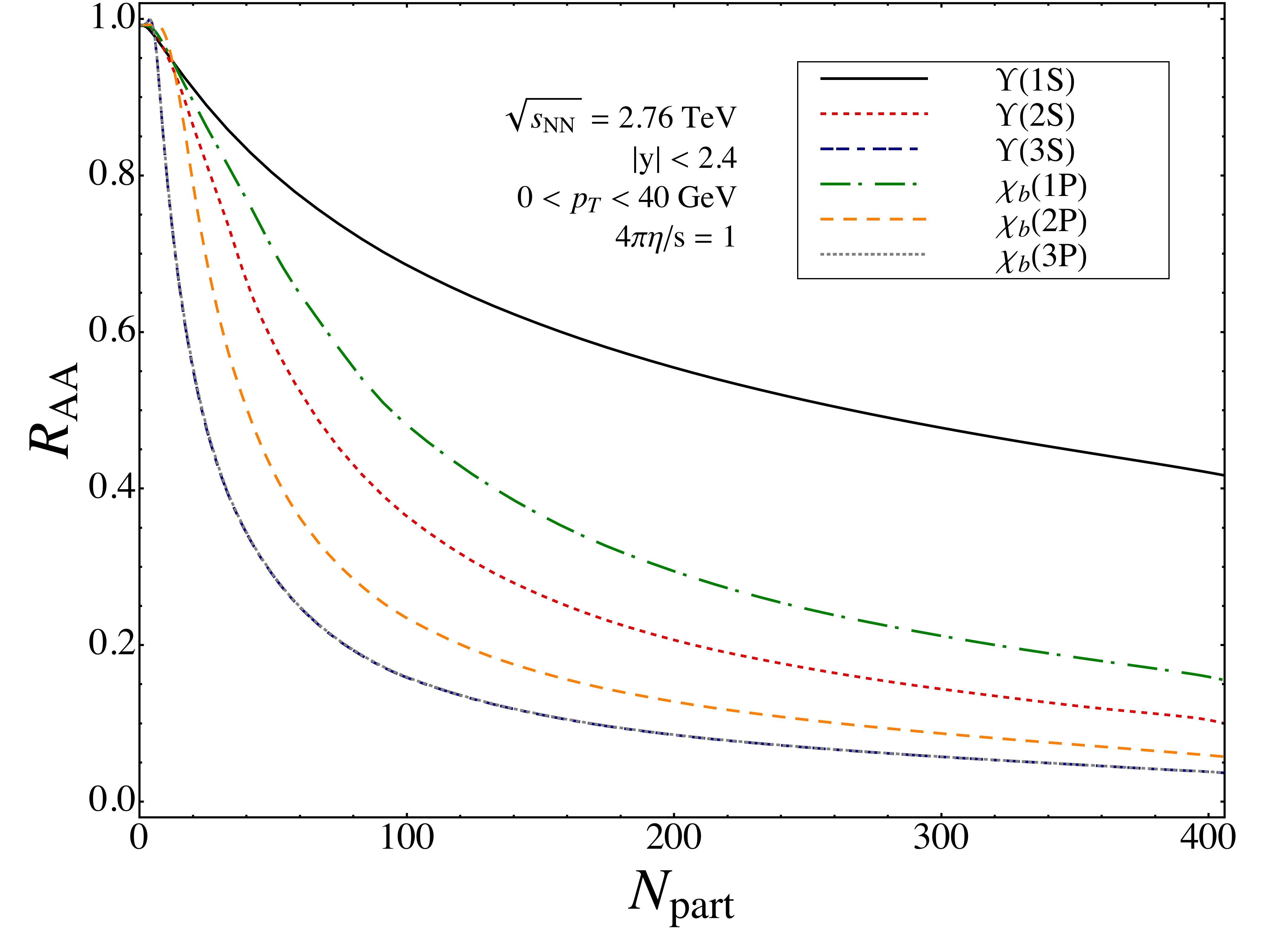}
\includegraphics[width=0.47\linewidth]{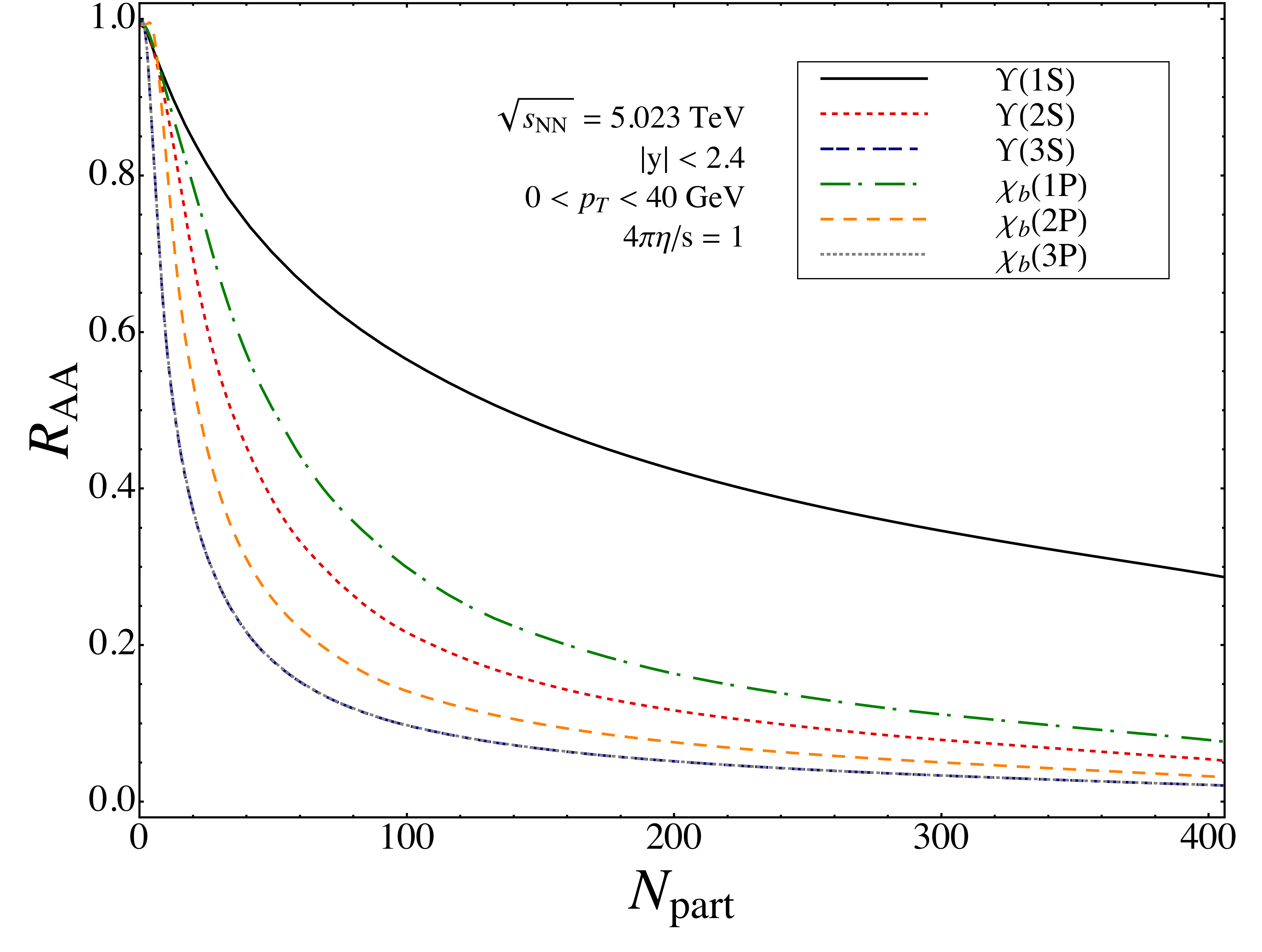}
}
\caption{
(Color online) Primordial $R_{AA}$ for each modeled state as a function of the number of participants. Note that the suppression curves for the $\Upsilon(3S)$ and $\chi_{b}(3P)$ states fall on top of each other.
}
\label{fig:rawcompare}
\end{figure}

We now turn to our results and predictions. In Figure \ref{fig:rawcompare}, we show the ``primordial'' $R_{AA}$ for the six states as a function of $N_{\text{part}}$ for the case of $4\pi\eta/s = 1$.  The primordial suppression is the result obtained prior to taking into account feed down effects.  The left panel shows the result for $\sqrt{s_{NN}} =$ 2.76 TeV, and the right panel shows our prediction for $\sqrt{s_{NN}} =$ 5.023 TeV.  From this figure, one can immediately see the sequential suppression of states, with excited states showing more suppression, which is a consequence of the lower dissociation temperatures compared to, for example, the $\Upsilon(1S)$ state.  Going~from 2.76 TeV to 5.023 TeV, we see increased primordial suppression for all states considered, with the ``primordial'' $R_{AA}$ decrease by approximately 31\%, 48\%, 44\%, 51\%, 45\%, and 44\% for the $\Upsilon(1S)$, $\Upsilon(2S)$, $\Upsilon(3S)$, $\chi_{b}(1P)$,
 $\chi_{b}(2P)$, and $\chi_{b}(3P)$ states, respectively, for central collisions. Finally, we note that at both energies, the model predicts significant primordial suppression of the $\Upsilon(1S)$, even though the temperatures probed in the plasma are below the state's disassociation temperature for most of the plasma evolution. The primordial suppression seen is a result of the fact that the decay rate remains large even below the naive disassociation temperature for a given state.

\begin{figure}[H]
\begin{center}
\includegraphics[width=0.6\linewidth]{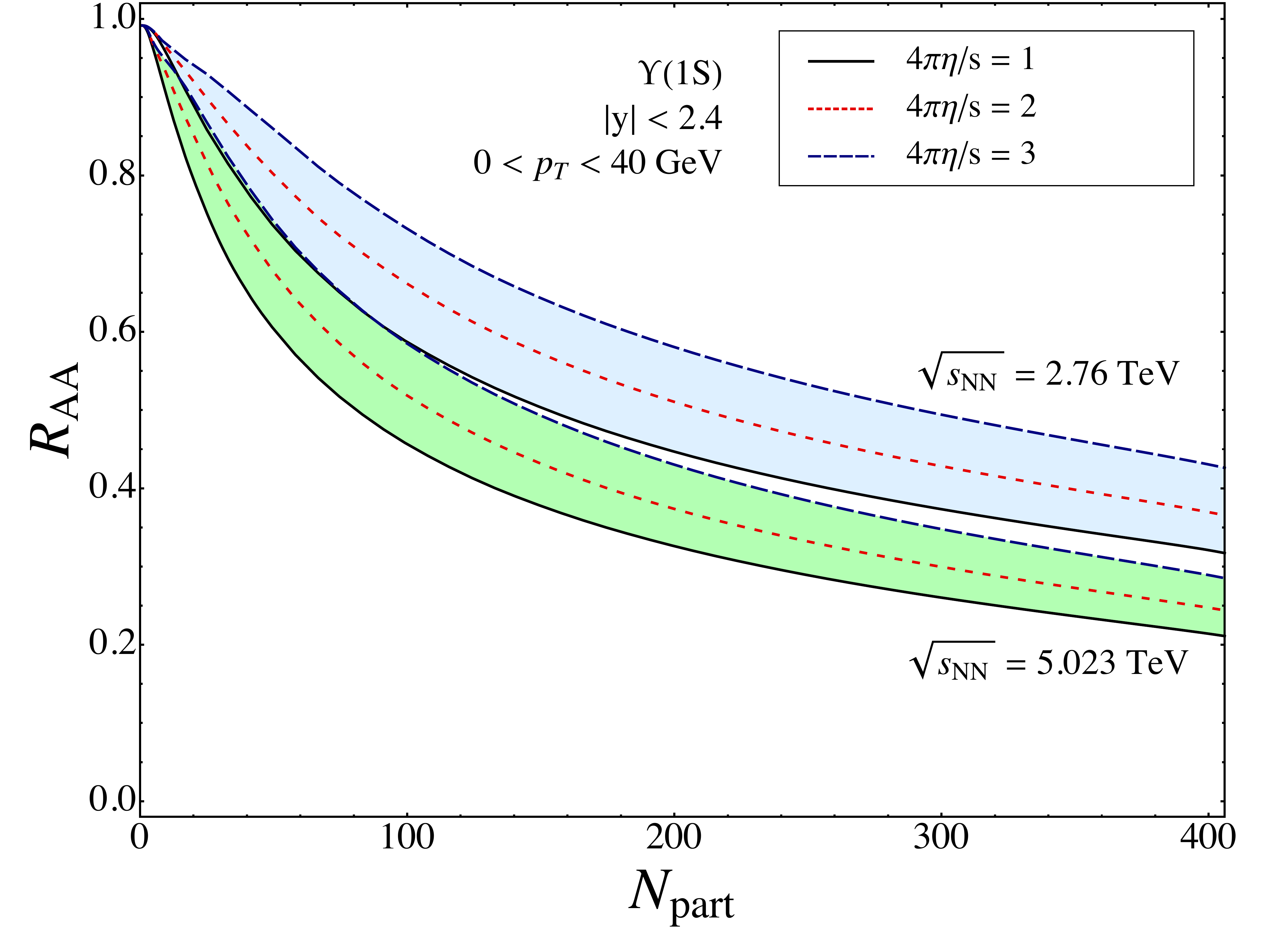}
\end{center}
\vspace{-8mm}
\caption{
(Color online) Inclusive $\Upsilon(1S)$ state calculated with feed down contributions from excited states. Here we show a comparison between $\sqrt{s_{NN}} = 2.76 \ \text{TeV}$ and $\sqrt{s_{NN}} = 5.023 \ \text{TeV}$ collision~energies.
}
\label{fig:npartfdcompare}
\end{figure}

In Figure \ref{fig:npartfdcompare}, we compare the inclusive suppression of the $\Upsilon(1S)$ state at 2.76 TeV and 5.023 TeV as a function of $N_{\text{part}}$.  The inclusive $R_{\rm AA}$ includes the effect of resonance feed down.  For~Figure~\ref{fig:npartfdcompare}, we integrated over rapidity in the range $|y|<2.4$ and transverse momentum in the range \mbox{$0 < p_T < 40$ GeV}.  The top band (light blue) corresponds to $\sqrt{s_{NN}} = 2.76$ TeV, and the bottom band (light green) corresponds to $\sqrt{s_{NN}} = 5.023$ TeV.  In each of the bands, the solid (black), short-dashed (red), and long-dashed (blue) lines correspond to $4\pi\eta/s \in \{1,2,3\}$.  As this figure demonstrates, our model prediction is that one should see enhanced suppression of the $\Upsilon(1S)$ at $\sqrt{s_{NN}} = 5.023$ TeV compared to $\sqrt{s_{NN}} = 2.76$ TeV.  For a central collision, we see an approximately 34\% decrease in the $\Upsilon(1S)$ $R_{\rm AA}$ for $4\pi\eta/s =1$.

\begin{figure}[H]
\begin{center}
\includegraphics[width=0.6\linewidth]{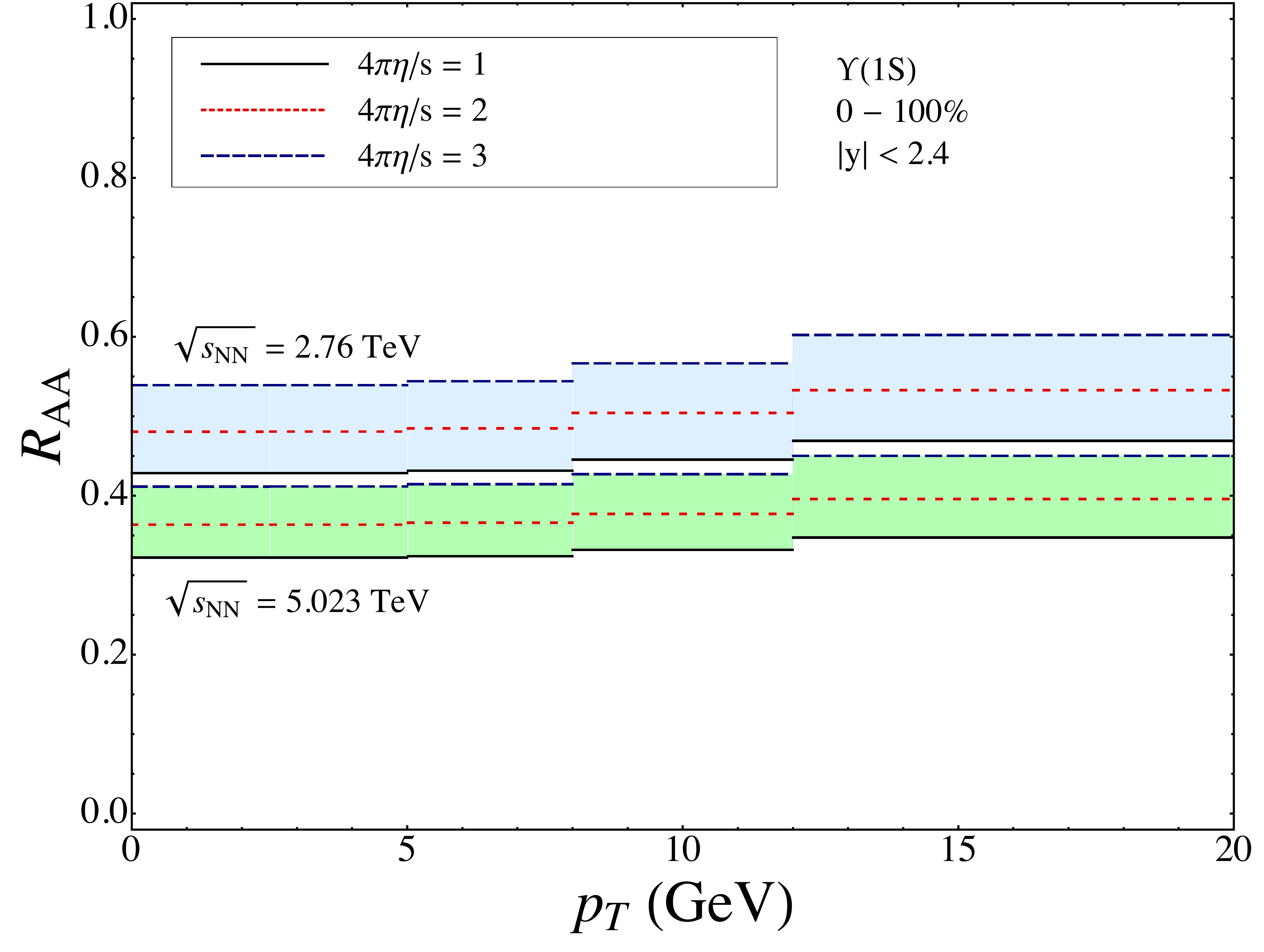}
\end{center}
\vspace{-6mm}
\caption{
(Color online) Inclusive $\Upsilon(1S)$ state calculated with feed down contributions from excited states. Here we show a comparison between $\sqrt{s_{NN}} = 2.76 \ \text{TeV}$ and $\sqrt{s_{NN}} = 5.023 \ \text{TeV}$ collision~energies.
}
\label{fig:ptfdcompare}
\end{figure}

\begin{figure}[H]
\begin{center}
\includegraphics[width=0.6\linewidth]{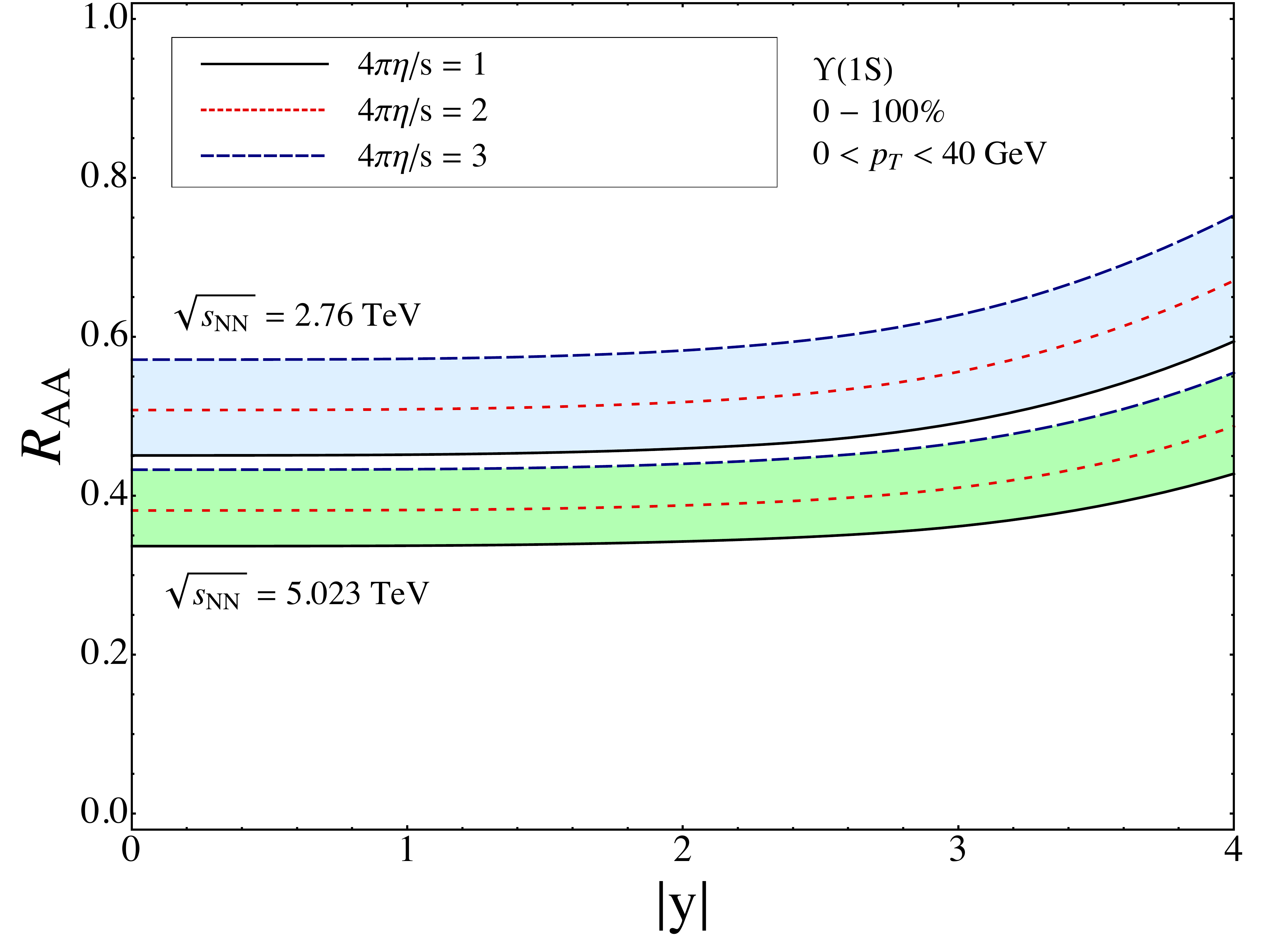}
\end{center}
\vspace{-6mm}
\caption{
(Color online)
 Inclusive $\Upsilon(1S)$ state calculated with feed down contributions from excited states. Here we show a comparison between $\sqrt{s_{NN}} = 2.76 \ \text{TeV}$ and $\sqrt{s_{NN}} = 5.023 \ \text{TeV}$ collision~energies.
}
\label{fig:rapfdcompare}
\end{figure}

In Figures~\ref{fig:ptfdcompare} and \ref{fig:rapfdcompare}, we compare the inclusive suppression of the $\Upsilon(1S)$ state at 2.76 TeV and 5.023 TeV as a function of $p_T$ and $y$, respectively.  The labelling and line types are the same as in Figure~\ref{fig:npartfdcompare}.  Due to the large size of the experimental $p_T$ bins, for Figure~\ref{fig:ptfdcompare} we have binned our predictions into the experimental bins.  From Figures~\ref{fig:ptfdcompare} and \ref{fig:rapfdcompare}, we find that the model once again predicts enhanced suppression of the $\Upsilon(1S)$ when going from 2.76 to 5.023 TeV as a function of $p_T$ and $y$.  In the lowest $p_T$ bin, we predict a decrease in $R_{AA}$ of approximately 25\% for the case $4\pi\eta/s =1$.  For $|y|=0$, we predict a decrease in $R_{AA}$ of approximately 26\% for the same case.  We also note that there is a slight increase in suppression for forward rapidities, which is due to the increased plateau halfwidth used in the initial conditions.

\begin{figure}[H]
\begin{center}
\includegraphics[width=0.6\linewidth]{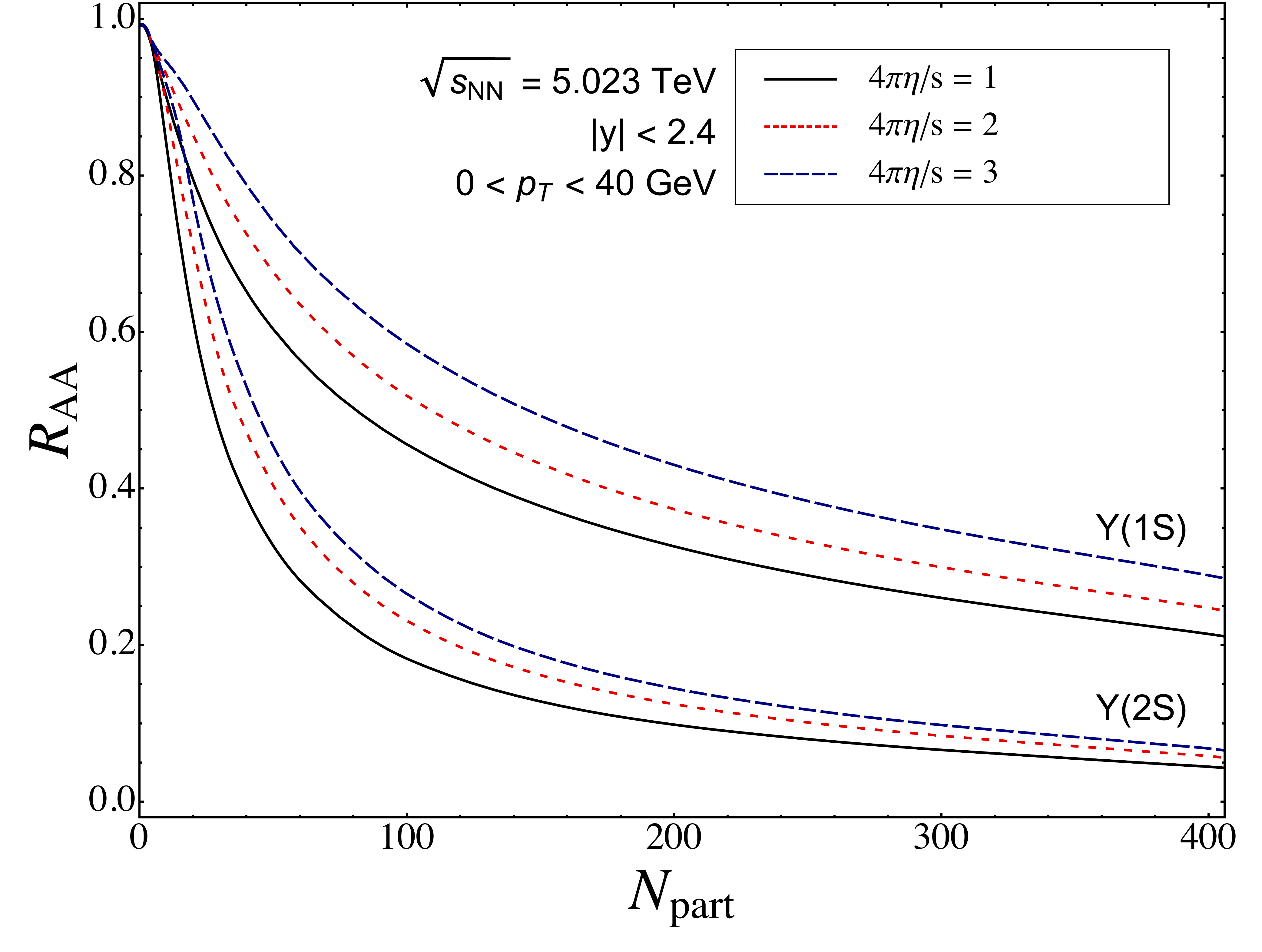}
\end{center}
\vspace{-7mm}
\caption{
(Color online) Predictions for inclusive $\Upsilon(1S)$ and $\Upsilon(2S)$ suppression for $\sqrt{s_{NN}} = 5.023 \ \text{TeV}$ Pb-Pb collisions.
}
\label{fig:npart5020}
\end{figure}

\begin{figure}[H]
\begin{center}
\includegraphics[width=0.6\linewidth]{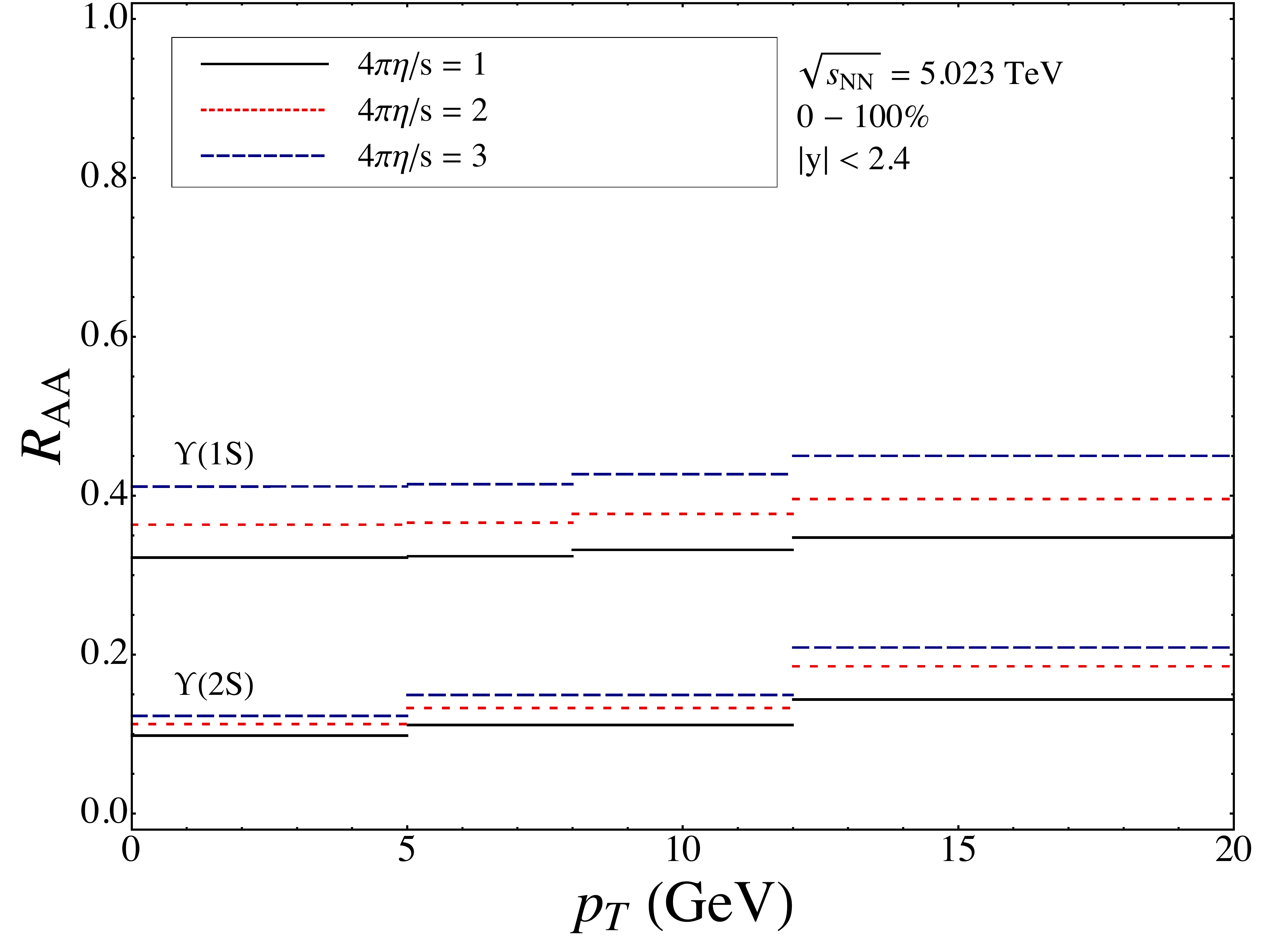}
\end{center}
\vspace{-6mm}
\caption{
(Color online) Predictions for inclusive $\Upsilon(1S)$ and $\Upsilon(2S)$ suppression for $\sqrt{s_{NN}} = 5.023 \ \text{TeV}$ Pb-Pb collisions.
}
\label{fig:pt5020}
\end{figure}

\begin{figure}[H]
\begin{center}
\includegraphics[width=0.6\linewidth]{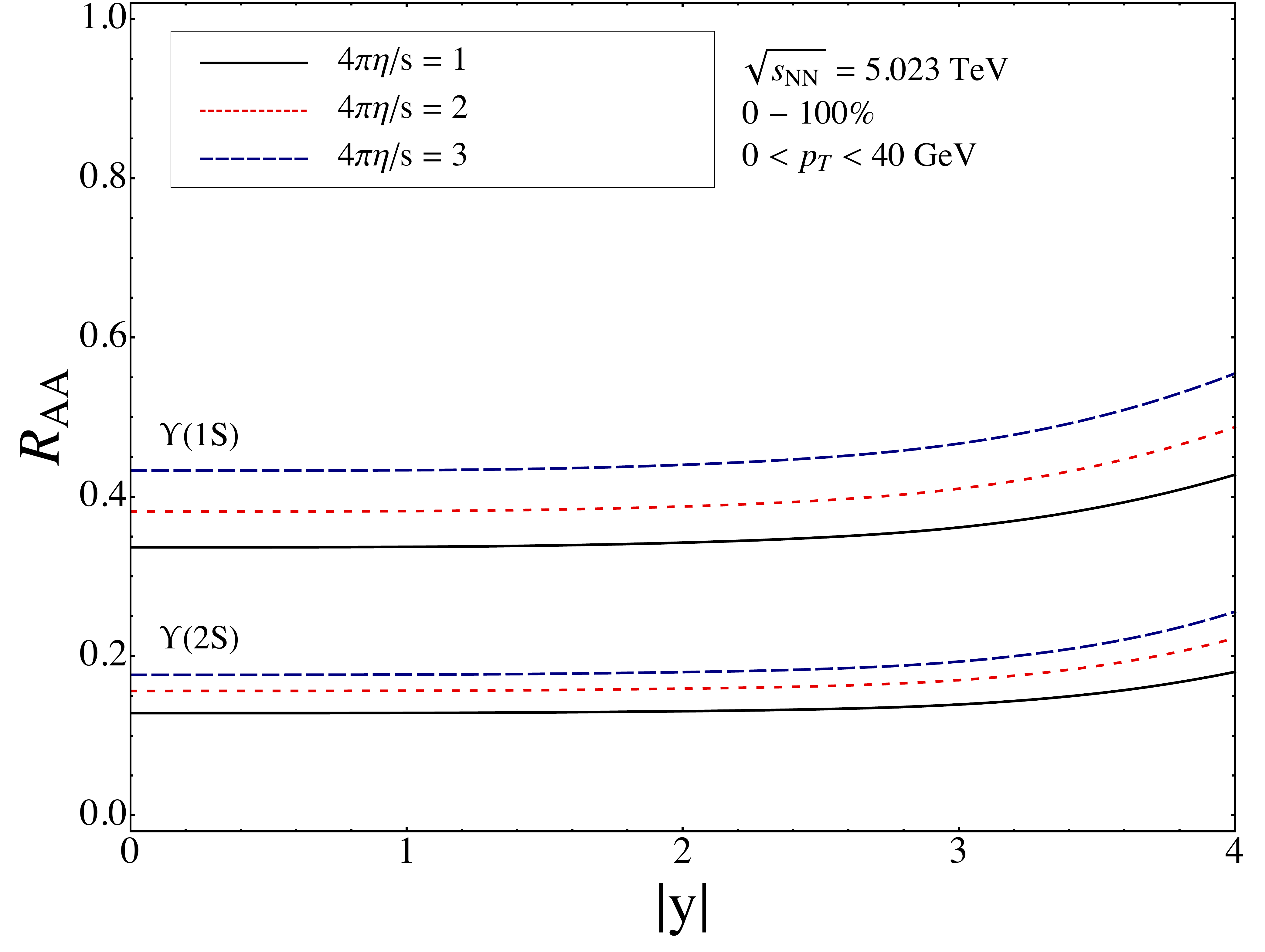}
\end{center}
\vspace{-6mm}
\caption{
(Color online) Predictions for inclusive $\Upsilon(1S)$ and $\Upsilon(2S)$ suppression for $\sqrt{s_{NN}} = 5.023 \ \text{TeV}$ Pb-Pb collisions.
}
\label{fig:rap5020}
\end{figure}

\begin{figure}[H]
\begin{center}
\includegraphics[width=0.6\linewidth]{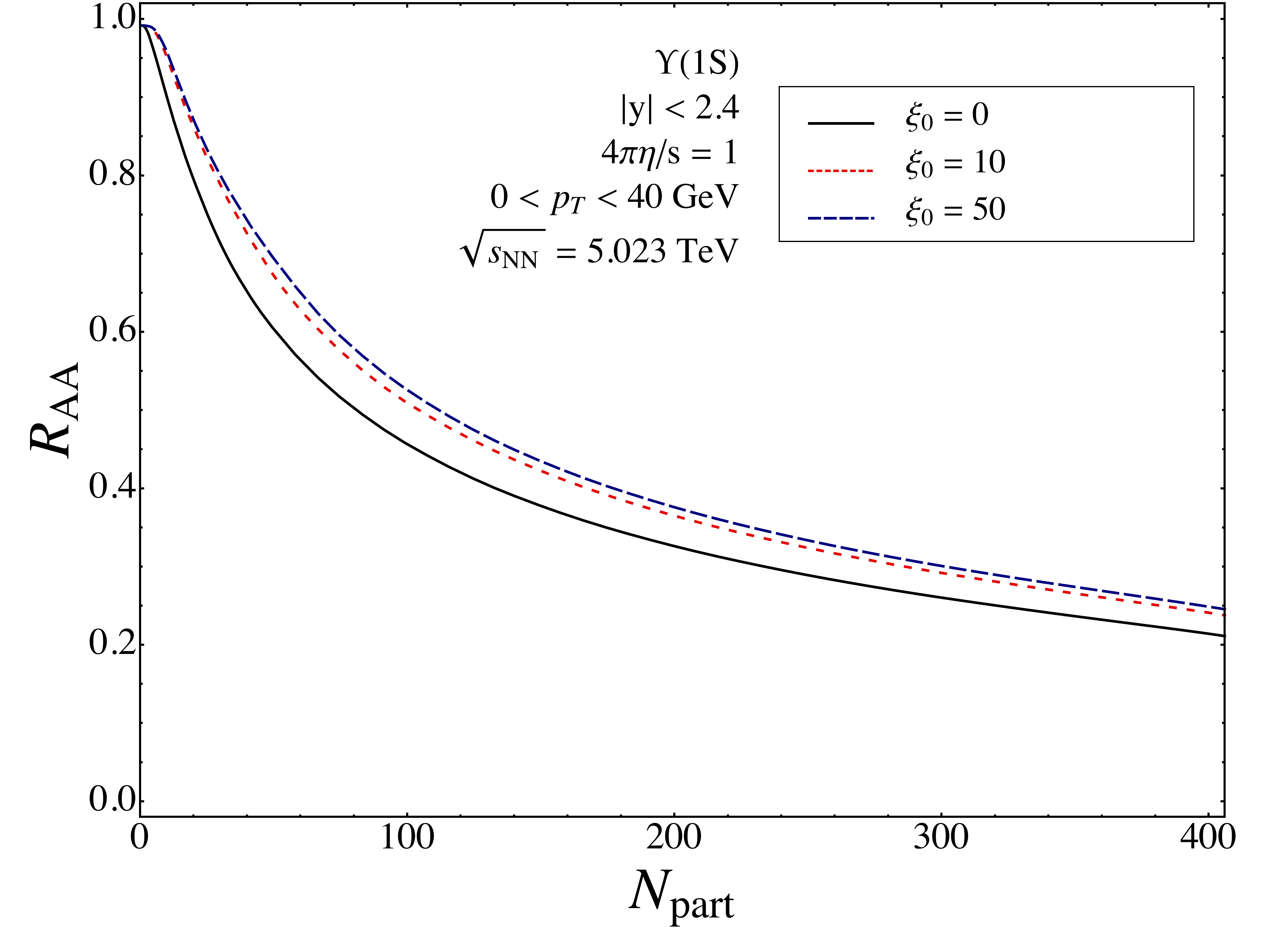}
\end{center}
\vspace{-8mm}
\caption{
(Color online) Inclusive $\Upsilon(1S)$ suppression for zero and finite initial anisotropy, $\xi_{0}$.
}
\label{fig:finitexi}
\end{figure}

In Figures \ref{fig:npart5020}--\ref{fig:rap5020}, we collect our predictions for the $\Upsilon(1S)$ and $\Upsilon(2S)$ inclusive $R_{AA}$ for \mbox{$\sqrt{s_{NN}} = 5.023 \ \text{TeV}$} collision energy.  The line styles are the same as in the previous figures.  As~can be seen from these figures, we also predict further suppression of the $\Upsilon(2S)$ at $\sqrt{s_{NN}} = 5.023 \ \text{TeV}$ as a function of $N_{\rm part}$, $p_T$, and $y$.  Finally, in Figure~\ref{fig:finitexi}, we plot the inclusive $\Upsilon(1S)$ suppression as a function of $N_{\rm part}$ for the case  $4\pi\eta/s =1$, but now varying the initial momentum-space anisotropy $\xi_0$ of the QGP.  The solid (black) line shows the case $\xi_0=0$, which corresponds to a QGP that is perfectly isotropic at $\tau_0$.  The short-dashed (red) line and long-dashed (blue) lines correspond to $\xi_0=10$ and $\xi_0=50$, respectively.  The finite values of $\xi_0$ map to initial pressure anisotropies in the local rest frame of ${\cal P}_L/{\cal P}_T = 0.13$ and $0.03$, respectively.  The presence of an initial momentum-space anisotropy is predicted by both weak and strong coupling approaches, with weak coupling approaches predicting larger initial momentum-space anisotropies than the strong coupling approaches~\cite{Strickland:2014pga,Fukushima:2016xgg}.


As Figure~\ref{fig:finitexi} demonstrates, the effect of initial momentum-space anisotropy is to decrease the amount of $\Upsilon(1S)$ suppression.  Going from $\xi_0$ = 0 to $\xi_0 = 50$, we see an approximately 14\% increase in the inclusive  $\Upsilon(1S)$ $R_{AA}$ for a central collision.  The effect of initial momentum-space anisotropy on the $\Upsilon(2S)$ is qualitatively similar, however, the effect is larger. We find that the inclusive $\Upsilon(2S)$ $R_{AA}$ increases by approximately 27\% for $\xi_0$ in the same range.

\section{Conclusions}

The primary goal of this work was to make predictions for bottomonia suppression at the latest LHC URHIC energy of $\sqrt{s_{NN}} = 5.023 \ \text{TeV}$.  In order to make our predictions, we had to make some extrapolations of the initial conditions for the background.  The key changes in going from 2.76 TeV to 5.023 TeV were a 16\% increase in the initial central temperature, an increase in the width of the central rapidity plateau~\cite{Ozonder:2013moa}, and a small change in the nucleon--nucleon scattering cross section~\cite{Miller:2007ri}.  With~these assumptions, we then made predictions with the same model employed in our prior works on this subject~\cite{Krouppa:2015yoa,Strickland:2011aa,Strickland:2011mw}.  The model includes the effect of in-medium dissociation of bottomonia, uses a full 3 + 1d anisotropic viscous hydrodynamics background, and takes into account non-equilibrium modifications of the heavy quark potential associated with large momentum-space anisotropy.  We did not include any cold nuclear matter effects or regeneration, since these are expected to be small for the bottomonia states.

Comparing higher energy 5.023 TeV collisions to 2.76 TeV collisions, our model predicts that one should see enhanced suppression of bottomonia states as a function of centrality, transverse momentum, and rapidity.  We have provided quantitative calculations of $R_{AA}$ for both the $\Upsilon(1s)$ and $\Upsilon(2s)$, including feed down effects which can be compared with forthcoming data from the LHC.

Finally, in addition to extrapolating our results to the higher energy, we looked at the effect of initial momentum-space anisotropy on bottomonia suppression.  For levels of momentum-space anisotropy predicted by theoretical models, we find an approximately 14\% increase in the $\Upsilon(1s)$ inclusive $R_{AA}$.  We plan to study this effect in more detail in the future, since it can affect the extracted value of $\eta/s$ when comparing with experimental data.

\vspace{6pt}
\acknowledgments{M. Strickland and B. Krouppa were supported by the U.S. Department of Energy under Award No. DE-SC0013470.}

\authorcontributions{M. Strickland and B. Krouppa contributed to both the theoretical and numerical aspects of the work presented herein.}

\conflictofinterests{{The authors declare no conflict of interest.}}
\bibliographystyle{mdpi}
\renewcommand\bibname{References}

\end{document}